\documentclass[%
 reprint,
 amsmath,amssymb,
 aps,
]{revtex4-1}

\usepackage{graphicx}
\usepackage{dcolumn}
\usepackage{bm}
\usepackage{hyperref}
\usepackage{longtable}

\makeatletter
\def\frontmatter@thefootnote{%
	\altaffilletter@sw{\@alph}{\@alph}\c@footnote
}%

\def\doauthor#1#2#3{%
	\ignorespaces#1\unskip\@listcomma
	\def\@tempifx{#3}%
	\@ifx{\@tempifx\@empty}{%
		\def\@tempifx{#2}%
		\@ifx{\@tempifx\@empty}{}{\frontmatter@footnote{#2}}%
	}{%
	\ifnum\c@affil=\@ne\relax\else#3\fi%
	\def\@tempifx{#2}%
	\@ifx{\@tempifx\@empty}{}{\ifnum\c@affil=\@ne\relax\else\comma@space\fi\frontmatter@footnote{#2}}%
}%
\space \@listand
}%
\def\@affil@script#1#2#3{%
	\@ifnum{#1=\z@}{}{%
		\par
		\begingroup
		\frontmatter@affiliationfont
		\ifnum\c@affil=\@ne\relax\else\textsuperscript{#1}\fi%
		#2%
		\@if@empty{#3}{}{\frontmatter@footnote{#3}}%
		\par
		\endgroup
	}%
}%
\makeatother

\begin{document}

\preprint{APS/123-QED}

\title{Closed analytical solutions of Bohr Hamiltonian with Manning-Rosen potential model}

\author{ {\bf M Chabab} $^a$ }%
\email[$^a$ ]{mchabab@uca.ma}
\author{ {\bf A Lahbas} $^b$}%
\email[$^b$ ]{alaaeddine.lahbas@edu.uca.ma}
\author{{\bf M Oulne} $^c$}
\email[$^c$ ]{oulne@uca.ma}

\affiliation{%
High Energy Physics and Astrophysics Laboratory, Faculty of Science Semlalia, Cadi Ayyad University,\\
P. O. B. 2390, Marrakech 40000, Morocco \\
}%


\begin{abstract}
In the present work, we have obtained closed analytical expressions for eigenvalues and eigenfunctions of the Bohr Hamiltonian with the Manning-Rosen potential for $\gamma-$unstable nuclei as well as exactly separable rotational ones with $\gamma\approx0$. Some heavy nuclei with known $\beta$ and $\gamma$ bandheads have been fitted by using two parameters in the $\gamma-$unstable case and three parameters in the axially symmetric prolate deformed one. A good agreement with experimental data has been achieved.

\end{abstract}

\pacs{Valid PACS appear here}

\maketitle


\section{\label{sec:level1}Introduction}
Nowadays several attempts have been devoted to construct analytical solutions of Schr\"{o}dinger equation associated with the Bohr Hamiltonian \cite{bohr} and various model potentials with the aim of describing appropriately shape phase transitions in atomic nuclei, particularly those which are related to the issue of critical point symmetries, namely : E(5) \cite{E5} and X(5) \cite{X5}. The first symmetry corresponds to transition from vibrational U(5) to $\gamma-$unstable nuclei O(6), while the second one corresponds to transition from vibrational spherical shape U(5) to prolate deformed nuclei SU(3). Besides, the symmetry E(5) \cite{E5} represents a solution of the Bohr Hamiltonian with a potential assumed to be independent of the $\gamma$ collective variable, while in the X(5) \cite{X5} symmetry case the $\gamma$ potential has a minimum at $\gamma=0$. In the framework of the interacting boson approximation (IBA) model \cite{Iachello1} the O(6) group has been first introduced to explain spectrum of the $\gamma$-unstable nucleus \cite{Iachello2}. The geometrical picture of the O(6) group has shown to emerge in the coherent state formalism of the IBM, giving similar $\gamma$-flat potential energy surface to the collective model \cite{Ginocchio}. Caprio and Iachello \cite{Iachello3} have extensively addressed the geometries of dynamical symmetries of IBM in relation to quantum phase transitions. Very recently, a more realistic calculation on the $\gamma$-softness has been presented within the IBM in connection to the microscopic potential energy surface \cite{Nomura}.

The aim of the present work is to search for adequate solutions of the eigenvalues and eigenvectors problem allowing better theoretical predictions for $\gamma$-unstable and axially deformed nuclei. For this purpose,  instead of the infinite square well potential used in the original works \cite{E5} and X(5) \cite{X5} we adopt the Manning-Rosen potential as used in \cite{mrosen} for the $\beta$-part of the nuclear potential. While, the $\gamma$-part is taken to be equal to the harmonic oscillator one in the case of prolate axially deformed nuclei. Thus, by means of the asymptotic iteration method \cite{iam1,iam2}, we derive closed analytical expressions for the energy spectrum and the corresponding wave functions. Similar works exist in the literature with other potentials like the Coulomb \cite{kratzer}, Kratzer \cite{kratzer}, harmonic oscillator \cite{harmonic}, Davidson \cite{davison} and Morse \cite{morse} potential.

 This paper is organized as follows : In Section II the asymptotic iteration method is briefly described. In section III, we achieved the exact separation of the Bohr Hamiltonian in $\gamma$-unstable and prolate axial rotor case.  Excited-state wave functions are given in Section IV, while the numerical calculations for energy spectra and comparisons with experimental data are presented in section V. Finally, Section VI contains our conclusion.

\section{ Overview of the asymptotic iteration method }
As already mentioned, the asymptotic iteration method (AIM) \cite{iam1,iam2} has been proposed and applied \cite{iam3,iam4,iam5,iam6,iam7,iam8} to solve the second-order homogeneous differential equation of the form
\begin{equation}
	y''(x)=\lambda_0(x)y'(x)+s_0(x)y(x)  \label{1}
\end{equation}
where the variables $s_0(x)$ and $\lambda_0(x)$ are sufficiently differentiable with $\lambda_0(x)\neq 0$.
The differential equation \eqref{1} has a general solution
\begin{align}
y(x)&=\exp\Big(-\int^{x}\alpha(x_1)dx_1\Big)\Big[C_2\nonumber \\+&C_1\int^{x}\exp\Big(\int^{x_1}[\lambda_0(x_2)+2\alpha(x_2)]dx_2
   \Big)dx_1\Big] \label{2}
\end{align}
If we have $n>1$, for sufficiently large n, $\alpha(x)$ values can be obtained
\begin{equation}
\frac{s_n(x)}{\lambda_n(x)}=\frac{s_{n-1}(x)}{\lambda_{n-1}(x)}=\alpha(x)
  \label{3}
\end{equation}
with
  \begin{align}
  \lambda_n(x)=&\lambda'_{n-1}(x)+s_{n-1}(x)+\lambda_0(x)\lambda_{n-1}(x)  \nonumber  \\
   s_n(x)=&s'_{n-1}(x)+s_0(x)\lambda_{n-1}(x),\ n=1,2,3....  \label{4}
  \end{align}
This method consists of converting the Schr\"{o}dinger equation into the form of Eq. \eqref{1} for a given potential. Then, $\lambda_n$'s and $s_n$'s parameters are computed by means of the recurrences relations of Eq. \eqref{4}. The energy eigenvalues are then calculated by means of the following quantization condition \cite{iam1}
\begin{align}
\Delta_n=s_n(x)\lambda_{n-1}(x)-\lambda_n(x)s_{n-1}(x)=0, && k=1,2,3,...
  \label{5}
\end{align}

\section{Exact separation of the Bohr Hamiltonian}
The original collective Bohr Hamiltonian is given by \cite{bohr}
\begin{multline}
    H=-\frac{\hbar ^2}{2B}$\Big[$ \frac{1}{\beta^4}\frac{\partial}{\partial\beta} {\beta^4}\frac{\partial}{\partial\beta}+ \frac{1}{\beta^2\sin3\gamma}\frac{1}{\partial\gamma}\sin3\gamma\frac{\partial}{\partial\gamma}-\\
  \frac{1}{4\beta^2}\sum_{k}\frac{Q_{k}^{2}}{\sin^2(\gamma-\frac{2}{3}\pi k)} \Big]+V(\beta,\gamma)
  \label{6}
\end{multline}
where $\beta$ and $\gamma$ are the usual collective coordinates,  $Q_k (k=1, 2, 3) $ are the components of angular momentum in the intrinsic frame, and $B$ is the mass parameter.
\subsection{The $\gamma-$unstable case}
For $\gamma-$unstable structure case, the potential energy is independent of $\gamma$, namely $V(\beta,\gamma)=V(\beta)$. The separation of variables is achieved by assuming a wave function of the form \cite{wilet} :
\begin{equation}
\Psi(\beta,\gamma,\theta_i)=\xi(\beta)\Phi(\gamma,\theta_i)
 \label{7}
\end{equation}
where $\theta_i(i=1,2,3)$ are the Euler angles. \\
Then the separation of the variables leads to a system of two differential equations, one containing only the $\beta$ variable, while the second contains the $\gamma$ variable and the Euler angles, $i.e.$ :
 \begin{equation}
  \left[ -\frac{1}{\beta^4}\frac{\partial}{\partial\beta} {\beta^4}\frac{\partial}{\partial\beta}+u(\beta)+\frac{\Lambda}{\beta^2}\right]\xi(\beta)=\epsilon \xi(\beta)  \label{8}
\end{equation}
\begin{multline}
 \left[- \frac{1}{\sin3\gamma}\frac{\partial}{\partial\gamma}\sin3\gamma\frac{\partial}{\partial\gamma}+
  \frac{1}{4}\sum_{k}\frac{Q_{k}^{2}}{\sin^2(\gamma-\frac{2}{3}\pi k)}\right]\Phi(\gamma,\theta_i)\\=\Lambda\Phi(\gamma,\theta_i)  \label{9}
\end{multline}
where the reduced energies and reduced potentials are defined as $\epsilon=2BE/\hbar^2$,  $u=2BV/\hbar^2$, respectively. $\Lambda$ is the separation constant.\\
The $\gamma$ and Euler angles equation \eqref{9} has been solved by B\`es\cite{bes}. In this equation, the eigenvalues of the second-order Casimir operator SO(5) are expressed in the following form $\Lambda= \tau(\tau + 3)$, where $\tau$ is the seniority quantum number, characterizing the irreducible representations of SO(5) and taking the values $\tau=0, 1, 2, ...$ \cite{rakavy}.

The values of angular momentum L occurring for each $\tau$ are provided by a well known algorithm and are listed in \cite{iachello}. The ground state band levels are determined by $L=2\tau$.

Concerning the radial part \eqref{8}, the Rosen-Manning potential $u(\beta)$ where a unit depth \cite{mrosen}  is used
\begin{equation}
u(\beta)=- \Big(1-\frac{e^{\beta_{0}\alpha}-1}{e^{\beta\alpha}-1}\Big)^2\label{10}
\end{equation}
where $\alpha$ is a screening parameter characterizing the range of the potential and the parameter $\beta_0$ indicates the position of the minimum of the potential. For small values of $\beta$, the Manning-Rosen potential reproduces exactly the Kratzer potential \cite{kratzer2} with an additive constant. Moreover, if we expand both potentials : Manning-Rosen and Kratzer around the minimum of the potential $\beta_0$, we get the harmonic-oscillator potential.

Through the following transformation of the radial function $\xi(\beta)=\beta^{-2}\chi(\beta)$, the radial equation \eqref{8} leads to  :
 \begin{equation}
  \left[ -\frac{\partial^2}{\partial\beta^2}+\frac{\Lambda+2}{\beta^2}-\Big(1-\frac{e^{\beta_{0}\alpha}-1}{e^{\beta\alpha}-1}\Big)^2\right]\chi(\beta)=\epsilon \chi(\beta)  \label{11}
\end{equation}
For $L$-wave ($L\neq$0 states), the above equation cannot be solved analytically. Then, in order to obtain quasi-analytical solutions of \eqref{11}, we need to employ Greene-Aldrich approximation \cite{greene} to the centrifugal term similar to other authors \cite{g1,g2,g3}, \begin{equation}
\frac{1}{\beta^2}\approx \alpha^2\frac{e^{-\alpha\beta}}{(e^{-\alpha\beta}-1)^2}  \label{12}
\end{equation}
This approximation is valid only for small values of the screening parameter $\alpha$.

To solve Eq. \eqref{11} within AIM \cite{iam1}, we consider the following ansatz for the function $f(y)$, with the new variable $y=e^{-\alpha\beta}$,
\begin{align}
 \chi(y)=y^\nu(1-y)^\mu f(y) \label{13}
 \end{align}
with
\begin{align}
\nu=\frac{\sqrt{1-\epsilon}}{\alpha}, && \mu=\frac{1}{2}+\sqrt{\frac{9}{4}+\frac{(e^{\alpha\beta_0}-1)^2}{\alpha^2}+\Lambda}
\label{14}
\end{align}
Substituting the approximation \eqref{12} and the function Eq.\eqref{13} into Eq.\eqref{11} we get
\begin{align}
f''(y)=&\frac{(1+2\mu+2\nu)y-(1+2\nu)}{y(1-y)}f'(y)\nonumber\\
+&\frac{\alpha^2(\nu+\mu)^2+\epsilon-e^{2\alpha\beta_0}}{\alpha^2y(1-y)}f(y)
\label{15}
\end{align}
The first and the second terms in the right hand side of the above equation represent the $s_0$ and $\lambda_0$ functions of Eq. \eqref{1} respectively. We compute the $s_n$ and $\lambda_n$ by the recurrence relations of equation \eqref{4}. Then, through the quantization condition of the method, given in Eq.\eqref{5} we obtain the generalized formula of the radial energy eigenvalues :
\begin{equation}
\epsilon _{n,\tau}=-\frac{1}{4}\left[\frac{\big(\alpha^2(\mu+n)^2-(A+2)^2\big)\big(\alpha^2(\mu+n)^2-A^2\big)}{\alpha^2(\mu+n)^2}\right]
\label{16}
\end{equation}
where $n$ is the principal quantum number and with:
\begin{align}
\mu=\frac{1}{2}+\sqrt{\frac{9}{4}+\Lambda+\frac{A^2}{\alpha^2}}, && A=e^{\alpha\beta_0}-1, && \Lambda= \tau(\tau + 3)
\label{17}
\end{align}

\subsection{The prolate axial rotor case}
In this case, the reduced potential $v(\beta,\gamma)=2BV/\hbar^2$ depends on the asymmetry $\gamma$. However, to achieve exact separation of variables, we assume a potential of the form \cite{kratzer,davison,morse,wilet}
\begin{equation}
 v(\beta, \gamma ) =u(\beta)+\frac{w(\gamma)}{\beta^2} \label{18}
\end{equation}
For the $\gamma$-potential, we use a harmonic oscillator \cite{X5,davison}
\begin{equation}
 w(\gamma ) =(3c)^2\gamma^2 \label{19}
\end{equation}
where $c$ is a free parameter.

As the $\gamma$ potential is minimal at $\gamma=0$, one can write the angular momentum term of Eq. \eqref{6} as \cite{X5}
 \begin{multline}
\sum_ {k=1,2,3}\frac{Q_{k}^{2}}{\sin^2(\gamma-\frac{2}{3}\pi k)} \\ \approx \frac{4}{3}(Q_1^2+Q_2^2+Q_3^2)+Q_3^2\left(\frac{1}{\sin^2\gamma}-\frac{4}{3}\right)
    \label{20}
\end{multline}
Using wave functions of the form
\begin{equation}
\Psi(\beta,\gamma,\theta_i)=F_L(\beta)\eta_K(\gamma)\mathcal{D}_{M,K}^L(\theta_i) \label{21}
\end{equation}
where $\mathcal{D}(\theta_i)$ denote Wigner functions of the Euler angles.
$L$  are the eigenvalues of angular momentum, while  $M$ and $K$ are the eigenvalues of the projections of angular momentum on the laboratory fixed $x$-axis and the body-fixed $x'$-axis respectively.

Then separation of variables leads to
\begin{equation}
   \Big[ -\frac{1}{\beta^4}\frac{\partial}{\partial\beta} {\beta^4}\frac{\partial}{\partial\beta}+\frac{\hat{\Lambda}+\frac{L(L+1)}{3}}{\beta^2}+u(\beta)\Big]F_L(\beta)=\epsilon F_L(\beta)  \label{22}
\end{equation}
\begin{multline}
   \Big[- \frac{1}{\sin3\gamma}\frac{\partial}{\partial\gamma}\sin3\gamma\frac{\partial}{\partial\gamma}+
  \frac{K^2}{4}\Big(\frac{1}{\sin^2\gamma}-\frac{4}{3}\Big) \\+w(\gamma) \Big]\eta_K(\gamma)={\hat\Lambda}\eta_K (\gamma) \label{23}
\end{multline}
where $\hat\Lambda$ represents a parameter coming from the exact separation of the variables, obtained from the $\gamma$ equation. The equation \eqref{23} has been solved \cite{X5} for the potential \eqref{19}  leading to

\begin{equation}
\hat\Lambda= (6c)(n_\gamma+1)-\frac{K^2}{3}  \label{24}
\end{equation}
where $n_\gamma$ is the quantum number related to $\gamma$-oscillations.

By using the Greene-Aldrich approximation \cite{greene} and after
solving the $\beta$ equation \eqref{22} with the Manning-Rosen potential \eqref{10} through AIM \cite{iam1}, we obtain for the energy spectrum the following formula
\begin{equation}
\epsilon _{n,L}=-\frac{1}{4}\left[\frac{\big(\alpha^2(\mu+n)^2-(A+2)^2\big)\big(\alpha^2(\mu+n)^2-A^2\big)}{\alpha^2(\mu+n)^2}\right] \label{25}
\end{equation}
where $n$ is the principal quantum number, and with
\begin{align}
\mu=\frac{1}{2}+\sqrt{\frac{9}{4}+\hat\Lambda+\frac{L(L+1)}{3}+\frac{A^2}{\alpha^2}}, && A=e^{\alpha\beta_0}-1, \nonumber \\ \hat\Lambda=(6c)(n_\gamma+1)-\frac{K^2}{3} \label{26}
\end{align}

\section{The radial wave functions}
The radial equations \eqref{11} and \eqref{22} have the same form. Then, the radial function $\xi(\beta)$ corresponding to the $n^{th}$ eigenvector of Eq.\eqref{15} becomes
\begin{align}
\xi(\beta)=\beta^{-2}\chi(\beta),&& y=e^{-\alpha\beta},&& \chi(y)=y^\nu(1-y)^\mu f(y) \\ \nonumber
 \label{27}
\end{align}
with  $\nu=\sqrt{1-\epsilon_{n,L}}/\alpha$ and $\mu$ in each case acquires the following form:
\begin{itemize}
\item For $\gamma$-unstable nuclei it is given by\\ $ \mu=\frac{1}{2}+\sqrt{\frac{9}{4}+\tau(\tau + 3)+\frac{A^2}{\alpha^2}}$
\item For axially symmetric prolate deformed nuclei it should be replaced by\\  $\mu=\frac{1}{2}+\sqrt{\frac{9}{4}+(6c)(n_\gamma+1)+\frac{L(L+1)-K^2}{3}+\frac{A^2}{\alpha^2}}$
 \end{itemize}
 The radial wave functions of Eq.\eqref{15} are obtained through Eq.\eqref{2}
 \begin{equation}
f(y)=N_{n,L \ 2 }F_1(-n,n+2\mu+2\nu;1+\nu;y)  \label{28}
\end{equation}
where $N_{n,L }$ is a normalization constant and $_2F_1$ are hyper-geometrical functions. In order to normalize the function $f(y)$, we convert hypergeometrical functions to Jacobi polynomials  by means of Eq. (4.22.1) given in \cite{hand}.  The radial wave function can be written as
  \begin{align}
\chi(t)=N_{n,L}(1-t)^{\nu}(1+t)^{\mu}P_{n}^{(2\nu,2\mu-1)}(t) && t=1-2y  \label{29}
\end{align}
Therefore, $N_{n,L}$ is obtained from the normalization condition:
 \begin{align}
 \int_0^{\infty} \xi^2(\beta)\beta^4 d\beta&= \int_0^{\infty} \chi^2(\beta) d\beta \nonumber \\&=\frac{1}{\alpha}\int_{-1}^{1} \frac{1}{1-t} \chi^2(t) dt=1 & \label{30}
 \end{align}
Using the usual orthogonality relation of Jacobi polynomials (see Eq. (7.391.1) of  \cite{hand2}) we get
\begin{multline}
N_{n,L}=$\Big($\frac{\mu+n}{2\alpha n!\nu(\nu+\mu+n)} $\Big)$^{-1/2}  \\ $\Big($ \frac{\big(\Gamma(2\nu+1)\Gamma(1+n)\big)^2}{\Gamma(2\nu+1+n)} \frac{\Gamma(2\mu+n)}{\Gamma(2\nu+2\mu+n)}$\Big)$^{-1/2}  \label{31}
\end{multline}
\section{Numerical results}
The obtained energy spectrum formulas Eq.\eqref{16} and Eq.\eqref{25} in both cases of $\gamma$-unstable and axially symmetric deformed nuclei, respectively are applied to calculate excitation energies of nuclei with a mass number $100\leq A \leq 250$. As it is shown above in $\gamma-$unstable case, the energy spectrum depends on two free parameters ($\alpha,\beta_0$) while in the axially symmetric prolate deformed nuclei it depends on three free parameters ($\alpha,\beta_0,c$). All these parameters have been fixed by fitting the excitation energies normalized to the energy of the first excited state $E(2^+_{g.s.})$. The root mean square (rms) deviation between the theoretical values and the experimental data

\begin{equation}
\sigma_{r.m.s.}=\sqrt{\frac{\sum_{i=1}^n(E_i(exp)-E_i(th))^2}{(n-1)E(2_1^+)^2}}
  \label{32}
\end{equation}
is used to evaluate the quality of our results. $n$ is the number of levels.

The excitation energies for 25 $\gamma$-unstable nuclei with mass number $A>100$ and $E(4^+_{g.s.})/E(2^+_{g.s.})<2.6$ are shown in Table \ref{table:mesh}. The fitted parameters come from the Manning-Rosen potential, namely : $\beta_0$ the minimum and $\alpha$ the screening  of the $\beta$ potential. The theoretical predictions are done with Eq.\eqref{16}  for low-lying bands which are classified by the principal quantum number $n$. The ground state band (g.s.) with $n=0$, the $\beta$-band with $n=1$ and the $\gamma$-band is obtained from degeneracies of the g.s. level.
From Table \ref{table:mesh}, it is clear that the Manning-Rosen potential gives generally good results in comparison with experimental data \cite{data}.

\begin{figure}
\includegraphics[scale=.88]{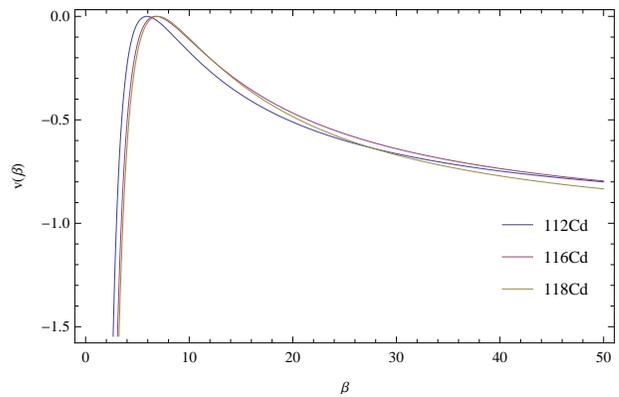}
\caption{ (Color online) Evolution of Manning-Rosen potential shapes for some $_{48}$Cd isotopes, with the
parameters of Table I. The quantities shown are dimensionless. } \label{fig1}
\end{figure}

The evolution and the shape  of Manning-Rosen potential for some $_{48}$Cd isotopes are shown in Fig.\ref{fig1} with the parameters given in Table \ref{table:mesh}. It is shown that the potential gets deeper as one moves from $^{112}_{48}$Cd to $^{118}_{48}$Cd, because $\beta_0$ and $\alpha$ increase (see Table \ref{table:mesh}).

\begin{figure}
\includegraphics[scale=.8]{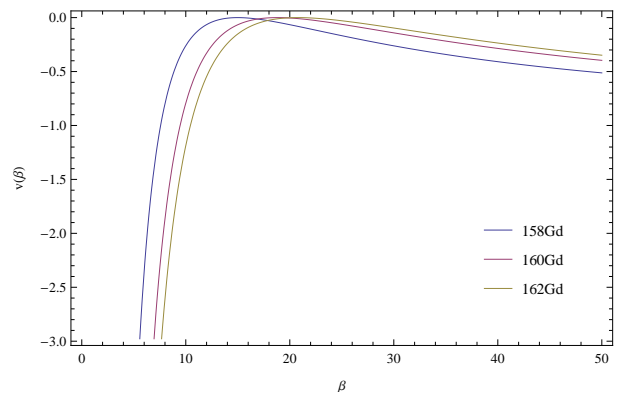}
\caption{ (Color online) Evolution of Manning-Rosen potential
shapes for some $_{64}$Gd isotopes, corresponding to the
parameters of Table II. The quantities shown are dimensionless. } \label{fig2}
\end{figure}
In Table \ref{table:mesh2},  we have also fitted the spectra of some axially symmetric deformed nuclei with mass number $A \geq 150$, $E(4^+_{g.s.})/E(2^+_{g.s.})>2.9$ for which the bandheads of the $\beta$ and $\gamma$ bands are known experimentally. The theoretical calculations are obtained from Eq.\eqref{25}, while the bands are classified by the  quantum numbers $n$, $n_{\gamma}$ and $K$, such as the ground state is $n=0$, $n_{\gamma}=0$, $K=0$, the $\beta$-band is $n=1$, $n_{\gamma}=0$, $K=0$, and the $\gamma$-band is $n=0$, $n_{\gamma}=1$, $K=2$. We remark a good agreement between our theoretical predictions and the experimental data in the g.s. and $\gamma$ bands, while in the $\beta$ band our results are relatively larger than the experimental ones.

In general the experimental ratios $R_{0/2}$ of the considered nuclei are well reproduced by the present model, while for a part of these nuclei it seems that other $\beta$ potentials as Davidson \cite{davison} or sextic \cite{Levai} type would be more appropriate for a better agreement.

The obtained Manning-Rosen potential  for some $_{64}$Gd isotopes is  shown in Fig. \ref{fig2} with the parameters given in Table \ref{table:mesh2}. As one moves from $^{158}_{64}$Gd to  $^{162}_{64}$Gd, the $\beta_0$ increases, but the $\alpha$ parameter with $c$ which is associated to the stiffness of the $\gamma$ potential, decrease.

In Fig.\ref{fig3} and Fig.\ref{fig4}, one can see that our calculations reproduce well the experimental energy spectra of the $^{166}$Er and $^{134}$Xe. A good agreement between our predictions and the experiment is achieved particularly for the g.s. and $\gamma$ bands. As to the $\beta$ band, a slight discrepancy between the theory and experiment is observed especially in the high levels region. Such a discrepancy may affect the quality factor $\sigma_{r.m.s.}$ of our theoretical predictions for some nuclei like $^{238}$U  (see Tables \ref{table:mesh} and \ref{table:mesh2} ). Despite this small defect in the $\beta$ band, as we can see in  Fig. \ref{fig5} the Manning-Rosen potential has proved its ability to reproduce adequately the experimental energy spectrum of $^{238}$U in comparison with the calculated spectrum with Davison potential within the deformation dependent mass formalism (DDMF) \cite{davi} bearing in mind that the DDMF \cite{davi, kra} introduces an extra fitting parameter which allows enhancing the precision of the numerical results, while our calculations have been carried out only with a constant mass parameter.

\section{Conclusion}
In this work,  we have investigated analytical solutions for Bohr Hamiltonian with Manning-Rosen potential. By using the improved Greene-Aldrich approximation for the
centrifugal term, we obtained the energy eigenvalues and the corresponding normalized radial wave functions expressed
in terms of the Jacobi polynomials.\\
 From mathematical point of view: this is achieved through the use of IAM.  Analytical expressions for the spectra and wave functions have been derived.
 From the physical point of view : spectra for $\gamma$-unstable and axially symmetric prolate deformed nuclei have been calculated and compared to experimental data. The obtained numerical results for more than 40 nuclei indicate that a good overall agreement with the experimental data is achieved in both cases.

\clearpage

\begin{table*}
\caption{{  \label{table:mesh} The comparison of theoretical predictions of the $\gamma$-unstable to experimental data (nuclei with $R_{4/2}<2.6$) for the ground state bandhead $R_{4/2}=E(4^+_{g.s.})/E(2^+_{g.s.})$ ratios, as well as those of the $\beta$ and $\gamma$ bandheads, normalized to the $2^+_{g.s.}$ state and labelled by $R_{0/2}=E(0^+_\beta)/E(2^+_{g.s.})$ and $R_{2/2}=E(2^+_\gamma)/E(2^+_{g.s.})$ respectively. $L_g$, $L_{\beta}$ and $L_{\gamma}$ characterized the angular momenta of the highest levels of the ground state, $\beta$ and $\gamma$ bands respectively, included in the fit.}}
\begin{tabular}{lllllllllllllll}
\hline

nucleus & $R_{4/2}$ & $R_{4/2}$ &$R_{0/2}$ & $R_{0/2}$ & $R_{2/2}$&$R_{2/2}$&$\beta_{0}$&$\alpha$&$L_g$&$L_\beta$&$L_\gamma$&$n$&$\sigma$\\
&exp&th&exp&th&exp&th\\
\hline

$ ^{112}$Pd & $2.533$ & $2.220$&$ 2.553$ & $2.871$&$2.113$ & $2.220$&$6.10$&$0.008$&$6$&$0$&$3$&$5$&$0.561$\\
$ ^{114}$Pd & $2.563$ & $2.390$&$ 2.622$ & $5.059$&$2.088$ & $2.390$&$10.43$&$0.01$&$16$&$0$&$11$&$18$&$0.988$\\
$ ^{106}$Cd & $2.361$ & $2.379$&$ 2.838$ & $2.872$&$2.713$ & $2.379$&$4.34$&$0.927$&$12$&$0$&$2$&$7$&$0.209$\\
$ ^{112}$Cd & $2.292$ & $2.203$&$ 1.983$ & $2.763$&$2.125$ & $2.203$&$5.91$&$0.005$&$12$&$8$&$11$&$20$&$0.575$\\
$ ^{116}$Cd & $2.375$ & $2.265$&$ 2.498$ & $3.203$&$2.362$ & $2.265$&$6.77$&$0.01$&$14$&$2$&$3$&$10$&$0.366$\\
$ ^{118}$Cd & $2.388$ & $2.281$&$ 2.636$ & $3.352$&$2.603$ & $2.281$&$6.96$&$0.02$&$14$&$0$&$3$&$9$&$0.417$\\
$ ^{128}$Xe & $2.333$ & $2.309$&$ 3.574$ & $3.892$&$2.189$ & $2.309$&$6.08$&$0.114$&$10$&$2$&$7$&$12$&$0.469$\\
$ ^{130}$Xe & $2.247$ & $2.312$&$ 3.346$ & $3.928$&$2.093$ & $2.312$&$6.13$&$0.113$&$14$&$0$&$5$&$11$&$0.563$\\
$ ^{132}$Xe & $2.157$ & $2.082$&$ 2.771$ & $2.824$&$1.944$ & $2.082$&$3.57$&$0.388$&$6$&$0$&$5$&$7$&$0.150$\\
$ ^{134}$Xe & $2.044$ & $1.894$&$ 1.932$ & $1.907$&$1.905$ & $1.894$&$2.99$&$0.232$&$6$&$0$&$5$&$7$&$0.105$\\

$ ^{130}$Ba & $2.524$ & $2.456$&$ 3.300$ & $3.317$&$2.541$ & $2.456$&$5.24$&$0.971$&$12$&$0$&$6$&$11$&$0.405$\\
$ ^{132}$Ba & $2.428$ & $2.357$&$ 3.237$ & $4.346$&$2.221$ & $2.357$&$8.70$&$0.02$&$14$&$0$&$8$&$14$&$0.846$\\

$ ^{134}$Ba & $2.316$ & $2.198$&$ 2.911$ & $2.851$&$1.931$ & $2.198$&$19.40$&$0.105$&$8$&$0$&$4$&$7$&$0.339$\\
$ ^{136}$Ba & $2.280$ & $2.062$&$ 1.929$ & $1.924$&$1.895$ & $2.062$&$3.26$&$0.795$&$6$&$0$&$2$&$4$&$0.159$\\
$ ^{142}$Ba & $2.322$ & $2.434$&$ 4.270$ & $4.293$&$3.960$ & $2.434$&$4.90$&$0.946$&$14$&$0$&$2$&$8$&$0.600$\\
$ ^{136}$Ce & $2.380$ & $2.273$&$1.949$ & $3.268$&$1.978$ & $2.273$&$6.97$&$0.007$&$16$&$0$&$3$&$10$&$0.717$\\
$ ^{138}$Ce & $2.316$ & $2.137$&$1.873$ & $2.426$&$1.915$ & $2.137$&$4.93$&$0.023$&$14$&$0$&$2$&$8$&$0.409$\\

$ ^{140}$Nd & $2.329$ & $2.021$&$1.827$ & $2.002$&$1.926$ & $2.022$&$4.07$&$0.012$&$6$&$0$&$2$&$4$&$0.230$\\

$ ^{140}$Sm & $2.347$ & $2.375$&$ 1.867$ & $1.890$&$2.676$ & $2.375$&$4.28$&$0.944$&$8$&$0$&$2$&$5$&$0.154$\\

$ ^{142}$Sm & $2.332$ & $2.247$&$ 1.888$ & $1.883$&$2.158$ & $2.247$&$3.69$&$0.887$&$8$&$0$&$2$&$5$&$0.169$\\
$ ^{192}$Pt & $2.479$ & $2.401$&$ 3.776$ & $3.764$&$1.935$ & $2.401$&$4.54$&$0.926$&$10$&$0$&$8$&$12$&$0.716$\\

$ ^{194}$Pt & $2.470$ & $2.343$&$ 3.858$ & $4.375$&$1.894$ & $2.343$&$6.80$&$0.102$&$10$&$4$&$5$&$11$&$0.803$\\

$ ^{196}$Pt & $2.465$ & $2.402$&$ 3.192$ & $3.908$&$1.936$ & $2.402$&$4.55$&$0.924$&$10$&$2$&$6$&$11$&$0.775$\\

$ ^{198}$Pt & $2.419$ & $2.160$&$ 2.246$ & $2.742$&$1.902$ & $2.160$&$4.33$&$0.160$&$6$&$2$&$4$&$7$&$0.489$\\

$ ^{200}$Pt & $2.347$ & $2.033$&$ 2.378$ & $2.373$&$1.845$ & $2.033$&$3.25$&$0.709$&$4$&$0$&$4$&$5$&$0.195$\\

\hline

\end{tabular}%


\end{table*}
\begin{table*}
\caption{\label{table:mesh2}  The comparison of theoretical predictions of the axially symmetric prolate deformed nuclei to experimental data (nuclei with $R_{4/2} > 2.9$) for the ground state bandhead $R_{4/2}=E(4^+_{g.s.})/E(2^+_{g.s.})$ ratios, as well as those of the $\beta$ and $\gamma$ bandheads, normalized to the $2^+_{g.s.}$ state and labelled by $R_{0/2}=E(0^+_\beta)/E(2^+_{g.s.})$ and $R_{2/2}=E(2^+_\gamma)/E(2^+_{g.s.})$ respectively. $L_g$, $L_{\beta}$ and $L_{\gamma}$ characterized the angular momenta of the highest levels of the ground state, $\beta$ and $\gamma$ bands respectively, included in the fit.} \normalsize

\begin{tabular}{lllllllllllllll}
\hline
nucleus & $R_{4/2}$ & $R_{4/2}$ &$R_{0/2}$ &$R_{0/2}$ & $R_{2/2}$& $R_{2/2}$&$\beta_{0}$&$\alpha$&$c$&$L_g$&$L_\beta$&$L_\gamma$&$n$&$\sigma$\\
&exp&th&exp&th&exp&th\\
\hline
$ ^{150}$Nd & $2.93$ & $3.14$&$ 5.19$ & $7.33$&$8.16$ & $8.58$&$6.70$&$0.006$&$3.5$&$14$&$6$&$4$&$13$&$1.086$\\

$ ^{152}$Sm & $3.01$ & $3.18$&$ 5.62$ & $8.35$&$8.92$ & $10.51$&$7.65$&$0.005$&$4.3$&$16$&$14$&$9$&$23$&$1.718$\\

$ ^{154}$Sm & $3.25$ & $3.27$&$ 13.41$ & $14.61$&$17.57$ & $18.88$&$14.03$&$0.003$&$7.2$&$16$&$6$&$7$&$17$&$1.056$\\


$ ^{158}$Gd & $3.29$ & $3.28$&$ 15.05$ & $15.22$&$14.93$ & $15.36$&$14.99$&$0.003$&$5.6$&$12$&$6$&$6$&$14$&$0.502$\\

$ ^{160}$Gd & $3.30$ & $3.29$&$ 17.62$ & $18.49$&$13.13$ & $13.26$&$18.82$&$0.001$&$4.6$&$16$&$4$&$8$&$17$&$0.965$\\

$ ^{162}$Gd & $3.29$ & $3.30$&$ 19.93$ & $20.33$&$11.98$ & $12.04$ & $20.78$ & $0.001$ &$4.1$&$14$&$0$&$4$&$10$&$0.298$\\

$ ^{162}$Dy & $3.29$ & $3.29$&$ 17.33$ & $17.92$&$11.01$ & $11.63$ & $18.17$ & $0.002$ &$4.0$&$18$&$8$&$14$&$26$&$1.310$\\
$ ^{164}$Dy & $3.30$ & $3.31$&$ 22.56$ & $22.82$&$10.38$ & $10.21$&$22.86$&$0.003$&$3.4$&$20$&$0$&$10$&$19$&$0.202$\\
$ ^{166}$Dy & $3.31$ & $3.28$&$ 15.00$ & $14.99$&$11.19$ & $11.38$&$14.84$&$0.005$&$4.0$&$6$&$2$&$5$&$8$&$0.139$\\
$ ^{166}$Er & $3.29$ & $3.29$&$ 18.12$ & $17.66$&$9.75$ & $10.01$&$16.97$&$0.009$&$3.4$&$16$&$12$&$14$&$26$&$0.337$\\
$^{178}$Yb & $3.31$ & $3.28$&$ 15.66$ & $15.77$&$14.54$ & $14.48$&$15.87$&$0.001$&$5.2$&$6$&$4$&$2$&$6$&$0.100$\\
$^{180}$W& $3.26$ & $3.28$&$ 14.64$ & $14.93$&$10.79$ & $11.36$&$14.99$&$0.003$&$4.0$&$24$&$0$&$7$&$18$&$0.837$\\
$^{184}$W& $3.27$ & $3.22$&$ 9.01$ & $9.86$&$8.12$ & $8.48$&$9.95$&$0.001$&$3.1$&$10$&$4$&$6$&$12$&$0.864$\\
$^{188}$Os& $3.08$ & $3.18$&$ 7.01$ & $8,37$&$4.08$ & $4.47$&$8.61$&$0.006$&$1.5$&$12$&$2$&$7$&$13$&$0.649$\\

$^{190}$Os& $2.93$ & $3.08$&$ 4.88$ & $6.18$&$2.99$ & $3.32$&$6.46$&$0.002$&$1.1$&$10$&$2$&$6$&$11$&$0.693$\\

$^{228}$Th& $3.23$ & $3.28$&$ 14.40$ & $15.04$&$16.78$ & $16.97$&$14.87$&$0.001$&$6.3$&$18$&$2$&$5$&$14$&$0.519$\\

$^{238}$U & $3.30$ & $3.31$&$ 20.64$ & $26.64$&$23.61$ & $24.58$&$26.64$&$0.001$&$8.6$&$30$&$4$&$27$&$43$&$1.892$\\

$^{248}$Cm & $3.31$ & $3.32$&$ 24.98$ & $27.58$&$24.17$ & $24.14$&$27.62$&$0.001$&$8.4$&$28$&$4$&$2$&$17$&$1.306$\\

$^{250}$Cf & $3.32$ & $3.31$&$ 27.02$ & $26.99$&$24.16$ & $24.08$&$25.68$&$0.005$&$8.4$&$8$&$2$&$4$&$8$&$0.082$\\

\hline
\end{tabular}
\end{table*}
\begingroup
\begin{figure*}[h]
\includegraphics[scale=1]{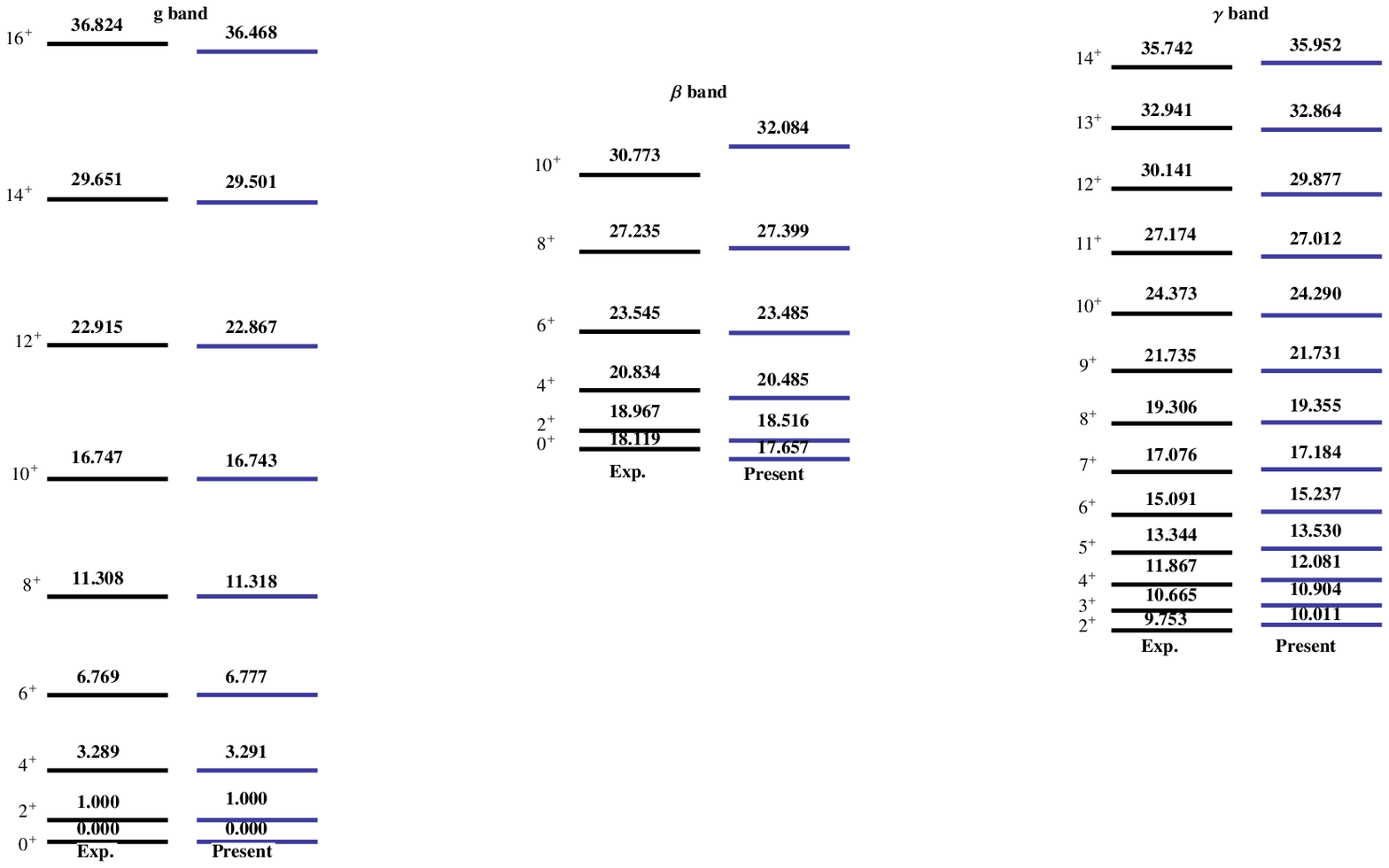}
\caption{ (Color online) The theoretical energy spectra, given by Eq. \eqref{25}, are compared with the  experimental data \cite{data} of the $^{166}$Er isotope, using the parameter sets given in Table \ref{table:mesh2}. }\label{fig3}
\end{figure*}
\endgroup
\begingroup
\begin{figure*}[h]
\includegraphics[scale=1]{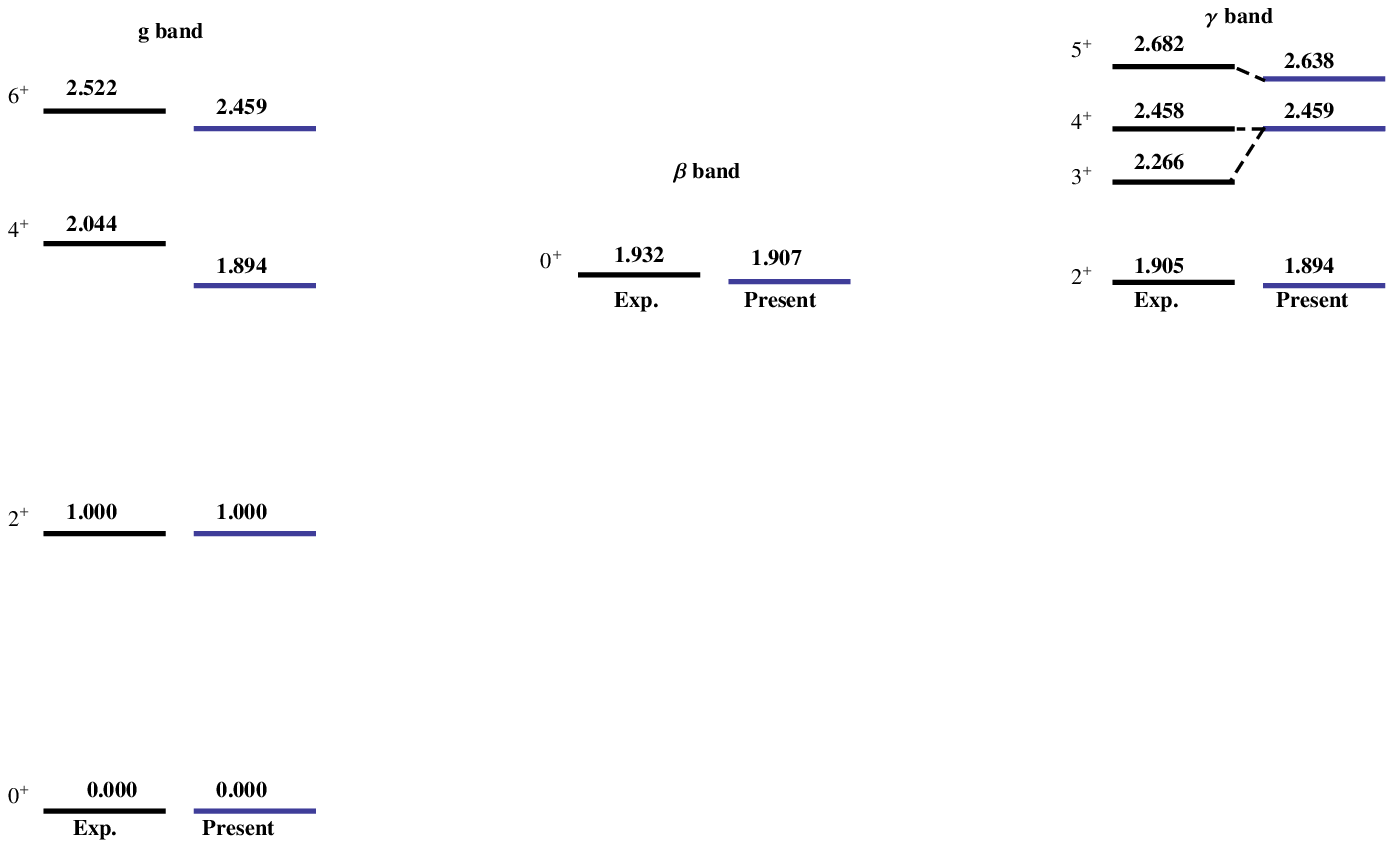}
\caption{(Color online) The theoretical energy spectra, given by Eq. \eqref{16}, are compared with the  experimental data \cite{data} of the $^{134}$Xe isotope, using the parameter sets given in Table \ref{table:mesh} } \label{fig4}
\end{figure*}
\begin{figure*}[h]
\includegraphics[scale=.85]{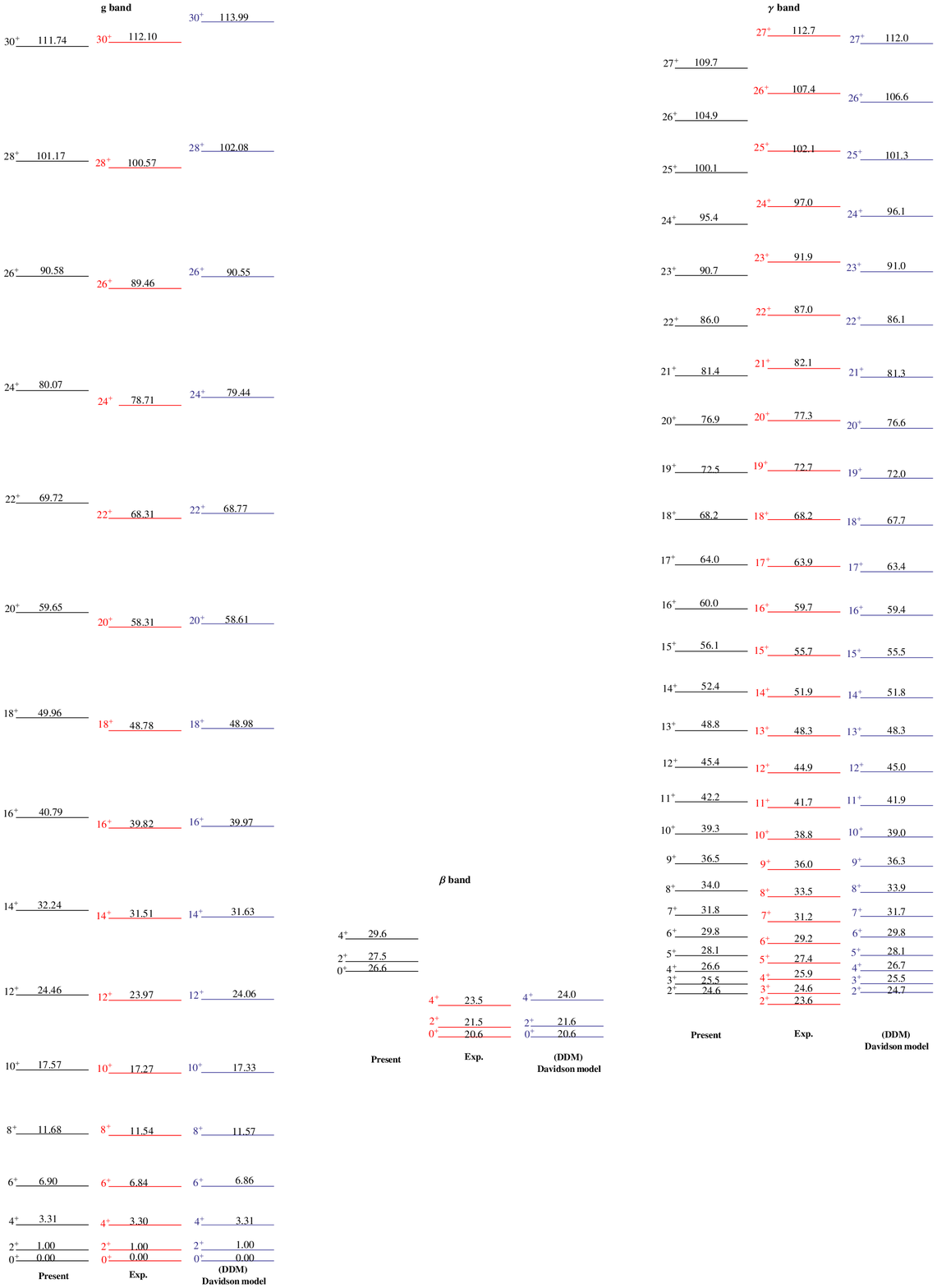}
\caption{(Color online) The theoretical energy spectra, given by Eq. \eqref{25}, are compared with the  experimental data \cite{data} of the $^{238}$U isotope and with the Deformation Dependent Mass (DDM) Davidson model \cite{davi}.} \label{fig5}
\end{figure*}
\endgroup


\begin{thebibliography}{}
\bibitem{bohr} A. Bohr, Mat. Fys. Medd. K. Dan. Vidensk. Selsk. 26, no. 14 (1952).
\bibitem{E5} F. Iachello, Phys. Rev. Lett. {\bf85}, 3580 (2000).
\bibitem{X5} F. Iachello, Phys. Rev. Lett. {\bf87}, 052502 (2001).
\bibitem{Iachello1} F. Iachello and A. Arima, {\it The Interacting Boson Model} (Cambridge University Press, Cambridge, 1987).
\bibitem{Iachello2} A. Arima and F. Iachello, Phys. Rev. Lett. {\bf40}, 385 (1978).
\bibitem{Ginocchio} J. N. Ginocchio and M. W. Kirson, Nucl. Phys. A {\bf350}, 31 (1980).
\bibitem{Iachello3} M. A. Caprio and F. Iachello, Ann. Phys. {\bf318}, 454 (2005).
\bibitem{Nomura} K. Nomura, N. Shimizu, D. Vretenar, T. Niksic, and T. Otsuka, Phys. Rev. Lett.
{\bf108}, 132501 (2012).
\bibitem{mrosen} Shi-Hai Dong and J Garc\'{i}a-Ravelo, Phys. Scr. {\bf75}, 307 (2007).
\bibitem{iam1} H. Ciftci, R. L. Hall, and N. Saad, J. Phys. A: Math. Gen. {\bf36}, 11807 (2003).
\bibitem{iam2} H. Ciftci, R. L. Hall, and N. Saad, J. Phys. A: Math. Gen. {\bf38}, 1147 (2005).

\bibitem{kratzer} L. Fortunato and A. Vitturi, J. Phys. G: Nucl. Part. Phys. {\bf30}, 627 (2004).
\bibitem{harmonic} D. Bonatsos, D. Lenis, E. A. McCutchan, D. Petrellis, and I. Yigitoglu, Phys. Lett. B {\bf649}, 394 (2007).
\bibitem{davison}D. Bonatsos, E. A. McCutchan, N. Minkov, R. F. Casten, P. Yotov, D. Lenis, D. Petrellis, and I. Yigitoglu, Phys. Rev. C {\bf76}, 064312 (2007).
\bibitem{morse}
I. Boztosun, D. Bonatsos, and I. Inci, Phys. Rev. C {\bf77}, 044302 (2008).
\bibitem{iam3}
I. Boztosun and M. Karakoc, Chin. Phys. Lett. {\bf24}, 3028 (2007).
\bibitem{iam4}
M. Chabab and M. Oulne, Int. Rev. Phys. {\bf4}, 331 (2010).
\bibitem{iam5}
M. Chabab, R. Jourdani, and M. Oulne, Int. J. Phys. Sci. {\bf7}, 1150 (2012).
\bibitem{iam6}
M. Chabab, A. Lahbas, and M. Oulne, Int. J. Mod. Phys. E {\bf21}, 10 (2012).
\bibitem{iam7}
M. Chabab, A. Lahbas, and M. Oulne, Phys. Rev. C {\bf91}, 064307 (2015).
\bibitem{iam8}
M. Chabab,  A. El Batoul, and M. Oulne J. Math. Phys. 56, 062111 (2015).
\bibitem{wilet}
L. Wilets and M. Jean, Phys. Rev.  {\bf102}, 788 (1956).
\bibitem{bes}
D. R. B\'es, Nucl. Phys. {\bf10}, 373 (1959).
\bibitem{rakavy}
G. Rakavy, Nucl. Phys. {\bf4}, 289 (1957).
\bibitem{iachello}
F. Iachello and A. Arima, The Interacting Boson Model (Cambridge University Press, Cambridge, 1987).
\bibitem{kratzer2}
A. Kratzer, Z. Phys. {\bf3}, 289 (1920).
\bibitem{greene}
R. L. Greene and C. Aldrich, Phys. Rev. A {\bf14}, 2363 (1976).
\bibitem{g1}
C. S. Jia, T. Chen, and L. G. Cui, Phys. Lett. A {\bf373}, 1621 (2009).
\bibitem{g2}
 S. H. Dong, W. C. Qiang, G. H. Sun, and V. B. Bezerra, J. Phys. A: Math. Theor. {\bf40}, 10535 (2007).
 \bibitem{g3}
 A. Soylu, O. Bayrak, and I. Boztosun, J. Phys. A: Math. Theor. {\bf41}, 065308 (2008).
 \bibitem{hand}
G. Szego, {\it Orthagonal Polynomials}, (American Mathematical Society, New York, 1939).\bibitem{hand2}
I. S. Gradshteyn and I. M. Ryzhik, Table of Integral, Series, and Products (Academic, New York, 1980).
\bibitem{Levai}
G. Levai, and J. M. Arias, Phys. Rev. C {\bf69}, 014304 (2004).
\bibitem{davi}
D. Bonatsos, P. E. Georgoudis, D. Lenis, N. Minkov, and C. Quesne, Phys. Rev. C {\bf83}, 044321 (2011).
\bibitem{kra}
D. Bonatsos, P. E. Georgoudis, D. Lenis, N. Minkov, D. Petrellis, and C. Quesne , Phys. Rev. C {\bf88}, 034316 (2013).

\bibitem{data}
http://www.nndc.bnl.gov/nndc/ensdf/.
\end{thebibliography}
\end{document}